\documentclass[aps,twocolumn,nofootinbib,preprintnumbers,citeautoscript]{revtex4}
\pdfoutput=1 
\usepackage{graphicx}
\usepackage{epsfig,amsmath,amsthm}
\usepackage{subfig}
\newcommand{\be}{\begin{equation}}
\newcommand{\ee}{\end{equation}}
\newcommand{\bea}{\begin{eqnarray}}
\newcommand{\eea}{\end{eqnarray}}
\def\bse{\begin{subequations}}
\def\ese{\end{subequations}}

\def\IZ{\relax\ifmmode\hbox{Z\kern-.4em Z}\else{Z\kern-.4em Z}\fi}

\def\bi{\begin{itemize}} \def\ei{\end{itemize}}

\def\({\left(} \def\){\right)}
\def\[{\left[} \def\]{\right]}

\begin{document}

\title{ \center{Compressible Matter at a Holographic Interface}}

\author{Moshe Rozali \\
{\it Department of Physics and Astronomy, \\
University of British Columbia,\\
Vancouver, BC V6T 1Z1, Canada} }

\begin{abstract}{We study the interface between a fractional topological insulator and an ordinary insulator, both described using holography. By turning on a chemical potential we induce a finite density of matter localized at the interface. These are gapless surface excitations which are expected to have a fermionic character. We study the thermodynamics of the system, finding a symmetry preserving compressible state at low temperatures, whose excitations exhibit hyperscaling violation. These results are consistent with the expectation of gapless fermionic excitations forming a Fermi surface at finite density.}
\end{abstract}

\maketitle

\section{Introduction}

Holography is a fairly recent approach to the study of strongly interacting systems which have gravitational duals. As such, it is ideal to model and gain qualitative understanding of otherwise  ill-understood phenomena in condensed matter systems. The approach has been applied successfully  to the study of homogeneous phases of holographic matter, and has provided insight into diverse phenomena such as superconductivity, quantum phase transitions, and non-Fermi liquid behavior.

Many interesting phenomena in modern condensed matter physics involve boundary excitations at the interface of two different bulk materials. Under suitable circumstances, those boundary excitations are gapless by virtue of topological considerations. Those considerations protect the excitations from developing a gap in the presence of noise and other small perturbations. These localized excitations are particularly interesting when they are predicted to be purely fermionic, as they are a natural candidate for an holographic non-Fermi liquid \cite{Karch}.

In this note we initiate the study of gapless surface excitations living on interfaces of different holographic materials. To that end, we consider the holographic model constructed in \cite{HoyosBadajoz:2010ac}  (for related development see \cite{Karch:2009sy,Maciejko:2010tx,Karch:2010mn,Sun:2011jc,Maciejko:2011ed,Swingle:2010rf}) which describes the interface of a {\it fractional} topological insulator and an ordinary (topologically trivial) insulator. Briefly, the model describes a fermion coupled to massless gauge fields, with spatially varying mass profile $m(x)$; index theorems then guarantee the existence of a fermionic bound state on the surface\footnote{The model also has charged bosonic excitations, but they are not expected to form bound states at the interface. We find below that there is no symmetry breaking in the model, i.e. the bosons are not condensed.}.
 Since the fermions are expected to be strongly correlated (by virtue of their coupling to deconfined gauge fields), it is  interesting to study the physics of these boundary excitations using various probes. 

Here we start such an investigation by studying the thermodynamics of the system. We find that the matter localized at the interface is {\it compressible} -- the density $d$ varies smoothly with the chemical potential $\mu$, with $d'(\mu) \neq 0$ for all values of the chemical potential (and $T \ll \mu$). Furthermore, the excitations around this state seem to exhibit a scaling symmetry with hyperscaling violation, within the precision of our numerics. These results suggest the existence of a Fermi surface for the boundary excitations.

The plan of this note is as follows: in section II we introduce the holographic setup and describe the inhomogeneous backgrounds we construct, where the surface excitations are put at finite density. In section III we discuss the thermodynamics of the system. We end by discussing directions for further exploration of the nature of these and other boundary excitations.

\section{The Holographic Setup}

We are interested in describing the surface excitations at non-zero temperature $T$ and finite chemical potential $\mu$. The starting point is then a black hole in  $AdS_5\times S^5$. We use the conventions of \cite{Mateos:2006nu,Kobayashi:2006sb,Mateos:2007vc}, work in the Poincare coordinates, and set the radius of curvature of $AdS_5$ to unity:
\bea
ds^2&=&\frac{1}{2} \pi^2 T^2 \left(-\frac{f^2(\rho)}{h(\rho)}dt^2+h(\rho)d\vec{x_3}^2\right)+ \frac{d\rho^2}{\rho^2}+d\Omega_5^2\nonumber \\
d\Omega_5^2 &=& d\theta^2+\sin^2 \theta\, d\Omega_3^2+\cos^2 \theta \,d\phi^2 \nonumber\\
f(\rho)&=& 1-\frac{1}{\rho^4} ~~~~~~~~~~~~~h(\rho)= 1+\frac{1}{\rho^4} \nonumber
\eea
We are interested in 7-branes embedded in $AdS_5$ and wrapping the 3-sphere inside $S^5$, localized at an angle $\theta$ on the five-sphere. The 7-brane embedding is characterized by the variation of $\theta$ as function of a radial coordinate $\rho$ in AdS, and in our case one of the boundary coordinates $x$.  The embedding is a minimal surface subject to asymptotic boundary conditions which encode the parameters of the problem: temperature, chemical potential and the mass parameter $m$. The precise definition of those asymptotic quantities appears below.

The embedding can end smoothly when the 3-sphere shrinks to zero size \cite{Karch:2002sh}, $\rho \,sin \theta=0$, which can happen in one of two ways:
\begin{itemize}

\item If $\rho \neq 0$ and $\sin \theta=0$, the 7-brane ends at a finite radial coordinate in AdS, away from the horizon, whose value depends (monotonically) on the mass parameter $m$.  Such embeddings are dubbed "Minkowski embeddings"; they correspond to states with no density\footnote{Minkowski embeddings with strings attached are possible in principle, but they give subleading contributions to the thermodynamics for any finite density.}. For zero temperature and $\mu < |m|$ these are the only possible embeddings and therefore in this parameter range the system is necessarily at zero density.

\item On the other hand, if $\rho = 0$, then $-1<\sin \theta<1$ parametrizes the inner boundary of the probe brane which lies on the black hole horizon. Such embeddings, called "black hole" embeddings  \cite{Kobayashi:2006sb,Mateos:2007vc}, can support finite density configurations and are possible, at zero temperature, only when $\mu>|m|$. 

\end{itemize}

In order to describe an interface we study the system with a mass profile $m(x)$ depending on a single boundary direction $x$, interpolating between $m(x)=-M$, a case which by convention we call the topological insulator, and $m(x)=M$ which corresponds in our sign conventions to a regular insulator. The system is insulating in the bulk, but supports gapless excitations on the interface (where $m(x)=0$). We are interested in probing the physics of those excitations. 

To that effect we turn on a constant\footnote{For a system in diffusive thermal equilibrium the chemical potential is constant even for inhomogeneous configurations.} chemical potential $\mu$. The effect of the chemical potential is easiest to understand at zero temperature. In that case, if we choose to satisfy $\mu<M$,  the chemical potential would induce no finite density in either one of the bulk insulators, but it would induce finite density at the interface, in the region where $m(x)<\mu$. The embedding then interpolates between the Minkowski embedding (ending at some finite radial coordinate), and the black hole embedding (ending at the black hole horizon). 

In practice, in order to avoid numerical instabilities associated with the abrupt transition between the two types of embeddings\footnote{Though it is not necessary for our purposes, it is interesting to construct solutions which correspond to phase boundaries in the holographic context, along the lines of \cite{Aharony:2005bm}.}, we choose the mass profile such that a small density is induced at the edges of the spatial interval. While the embedding is then of the "black hole" type throughout the spatial interval, the quantities calculated are largely independent of the mass profile, as long as the density induced is at least moderately peaked at the interface. 



\subsection*{Ansatz and Equations of Motion}

To describe the embedding we choose to parametrize the worldvolume in terms of the radial coordinate  $\rho$ and the spatial coordinate $x$. In our parametrization, the embedding is characterized by a single function $\chi(\rho,x)= \cos  \theta(\rho,x)$.  The asymptotic behavior of the coordinate $\chi$ at large $\rho$, $\chi(\rho,x)\rightarrow \frac{m(x)}{\rho}$, determines the mass profile $m(x)$, we choose to approximate a step function as $m(x) =\frac{2 M}{1+ e^{-a x} }-M $ 
for $a \gg 1$. As long as the mass profile is sufficiently narrow and steep, the parameters $a,M$ do not significantly influence the physical quantities we calculate\footnote{We checked that changing those parameters by two orders of magnitude changes the calculated quantities by less than one part in $10^4$.}. 

Having chosen a parametrization,  we add a worldvolume gauge field  $ A_0$ (choosing $2\pi \alpha'=1$). It is straightforward to find the following DBI action:
\bea
S_{DBI}&=& \mathcal{N}\, T^4\,\rho^3 \,(1-\chi^2)\,f \, h\, \sqrt {I_\chi+I_{A_0}+ I_{int}}  \nonumber \\
I_\chi &=&  1-\chi^2 + (\partial_\rho \chi)^2  +\frac{(\partial_x \chi)^2}{ \rho^2 \, h}  \nonumber \\
I_{A_0} &=& (1-\chi^2) \left(1 - \frac{2 h (\partial_\rho A_0)^2}{f^2}- \frac{2 (\partial_x A_0)^2}{\rho^4 \, f^2} \right)\nonumber \\
I_{int}&=& -\rho^2 (\partial_\rho \chi \partial_x A_0- \partial_\rho A_0\partial_x \chi )^2\nonumber
\eea
where the form of $I_\chi$ and $I_{A_0}$ reveals the elliptic nature of the equations. The field $A_0$ and the coordinate $x$ are chosen to be dimensionless, like all other quantities appearing in the action. With this choice we measure dimensionful quantities in units of the temperature $T$. Therefore, the asymptotic value of $A_0$ at large $\rho$ determines the chemical potential in units of $T$,  $A_0\rightarrow \tilde{\mu}= \frac{\mu}{T}$. 

The regularity of the solution near the boundary of the embedding  is discussed in \cite{Frolov:2006tc}. For a black hole embedding we need to satisfy $\partial_\rho \chi=0$ and $A_0=0$ at $\rho=0$. Asymptotically in the $\rho$ direction, we set a UV cutoff $\rho=\Lambda$ and choose  $A_0(\rho=\Lambda,x) =\mu$ to set the chemical potential, and  $\chi(\rho=\Lambda,x) = \frac{m(x)}{\rho}$ to specify the mass profile. Finally, we choose homogeneous Neumann boundary conditions for both functions at the boundaries of the spatial interval.

\begin{figure}[t]
   \centering
     \vspace{-60pt}
    \includegraphics[width=0.45 \textwidth]{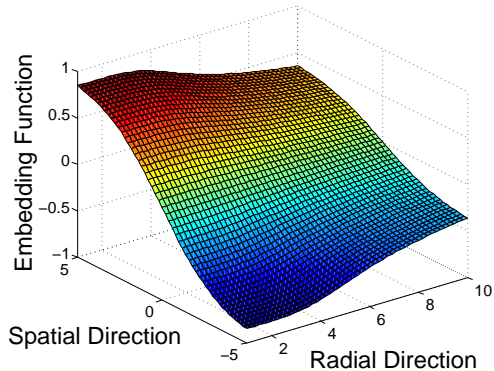}
      \vspace{-80pt}
    \caption{The embedding function $\chi$ near the horizon. The physical parameter is $\mu/T=2$. 
}
\end{figure}
\begin{figure}[b]
   \centering
    \vspace{-80pt}
    \includegraphics[width=0.55 \textwidth]{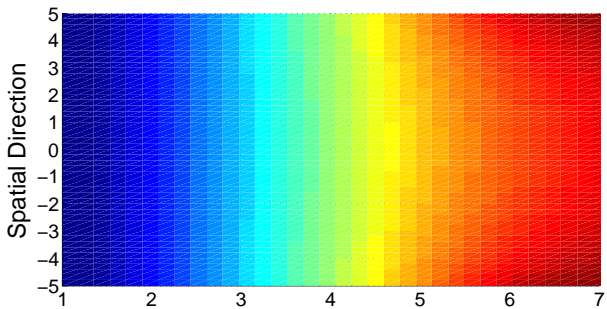}
     \vspace{-140pt}
    \caption{Contour plot of the gauge potential $A_0$ that shows the spatial inhomogeneity. The horizontal direction is a section of the the radial coordinate. The physical parameter is $\mu/T=3.4$. }
\end{figure}

\subsection*{The Solutions}
We solve the equations of motion numerically using pseudospectral methods implemented in Matlab, utilizing a Chebyshev grid to discretize the equations, and solving the resulting set of algebraic equations using a Newton-Raphson-Kantorovich iteration. The convergence of the solution with the linear size of the grid is exponential. The results shown are for a grid of 900 points (giving accuracy of about one part in $10^6$ for the values of the fields).

Before extracting physical properties, we briefly discuss the features of the solutions. In figures (1-3) we display a few characteristic solutions, at various values of the parameters. 

In figure 1 we display the embedding function $\chi$. The spatial variation of $\chi$ increases towards the horizon, where its profile is correlated with the boundary density: the density is maximal for $\chi=0$ and decreases to zero as $\chi\rightarrow \pm 1$. The value
$\chi = \pm 1$ at the horizon would correspond to the transition to the Minkowski embedding which carries no density.
 
In figure 2 we show a contour plot of the bulk gauge field. The spatial variation is small around the horizon and grows towards the boundary. Note that the leading behavior of the gauge field near the boundary is spatially homogeneous, and any inhomogeneity is manifested in the behavior of the subleading term, related to the density.

\vspace{-10pt}
 \section{Thermodynamics}
 
The thermodynamics of the system is completely specified when expressing the number density $n$ in terms of the chemical potential $\mu$ and the temperature $T$. Equivalently one can specify the free energy $F(\mu,T)$, then $n= \frac{\partial F}{\partial \mu}$. We choose to calculate the density $n$, which is read off the subleading fall-off of the gauge field, $A_0\rightarrow \tilde{\mu} -\frac{T^2}{\rho^2}\,n(\tilde{\mu})$ as $\rho\rightarrow \infty$, where $\tilde{\mu}=\frac{\mu}{T}$. 

The subleading term in the expansion of $\chi$, related to the chiral condensate in e.g. \cite{Kobayashi:2006sb}, is found to vanish at the interface for all our solutions. This is a result of a parity symmetry preserved by our mass profile, and indicates the absence of symmetry breaking and the existence of non-trivial fluid at the interface. The physics of that fluid depends only on $mu$ and $T$, leading to the dimensional analysis below.

Since our background is inhomogeneous, there is separate information in densities per unit volume (at fixed spatial position $x$), and the corresponding densities per unit area (when integrating over the spatial direction $x$). Here we mainly discuss the free energy and number densities {\it per unit volume}, at the center of the interface $x=0$, ending with comments on the integrated quantities.

Based on dimensional analysis in 3+1 dimensions, $F=T^4 \, \mathcal{F} (\tilde{\mu})$, since $T,\mu$ are the only scales in the problem \cite{Nogueira:2011sx}.  
For relativistic systems which form a Fermi surface at low temperatures, for example for free massless fermions\footnote{The normalization of this expression depends on the normalization of the fields in the microscopic theory, for example for us this would  involve the normalization of the fermion bound state wavefunction.}, $\mathcal{F}(\tilde{\mu}) \sim \tilde{\mu}^4+...$.  The leading term in this expansion corresponds to $n\sim \mu^3$, the contribution to the density from a putative Fermi surface.

The subsequent terms in the small $\tilde{\mu}$ expansion give information on low energy excitations of the system. In particular, the second term in the expansion gives the leading temperature dependence of the free energy at low temperatures. Thus, this term is responsible for the temperature scaling of the entropy $S$ in that regime. The entropy scales differently for systems with or without a Fermi surface. For degenerate relativistic fermions, one has $S_f \sim T$ at low temperatures, whereas for relativistic bosons  $S_b \sim T^3$ (see for example \cite{thermal}). The larger entropy at low temperatures reflect the fact that systems with a Fermi surface have more low energy excitations in the degenerate limit\footnote{Note that the scaling of the entropy depends only the density of low energy states and does not assume those excitations are weakly interacting.}. 

The difference between bosons and fermions can also be parametrized in terms of the  {\it hyperscaling violation} exponent $\theta$, such that $ S \sim T^{3-\theta}$ with $\theta=0$ for bosons and $\theta=2$ for fermions. The origin of the difference in scaling is the nature of the excitations of the Fermi surface. The low energy excitations are constrained to be close to the Fermi surface, thus reducing the effective dimensionality of the system.

After parametrizing the possible small temperature behavior, we can present the results of our model. Figure (4) displays a log-log plot of the density as function of the chemical potential (both measured in units of temperature). The small temperature expansion discussed in this section corresponds to the large $\tilde{\mu}$ limit. The functional dependence encoded  in figure (4) reveals a few interesting features of the system studied.

The most robust feature is the leading behavior at large $\tilde{\mu}$, $n(\tilde{\mu})\sim \tilde{\mu}^\alpha$, with $\alpha \simeq 3$ to numerical accuracy. Depending on fit options, both for reading off the density for each solutions and for fitting the curve $n(\tilde{\mu})$, $\alpha$ is fairly robust, changing between 2.95 and 3.1 (for the displayed graph $\alpha= 2.98$). 

This result is consistent with the expectation that the ground state of the system is a Fermi surface.
This relation implies that the derivative of the charge density $d$ with respect to the chemical potential is non-zero at low temperatures, i.e. the state at the  interface is {\it compressible}.  A Low temperature compressible state, in the absence of symmetry breaking, can be taken as indicative of a Fermi surface (see e.g. the discussion in \cite{Huijse:2011hp}).

Reading off the subleading terms in the $\tilde{\mu}$ expansion is less robust. Nevertheless, one can develop an asymptotic expansion in $\frac{1}{\tilde{\mu}}$, a low temperature expansion. To probe nature of the low energy excitations, parametrized by the value of $\theta$, we fit the  asymptotic data to the form\footnote{A quadratic term would correspond to $\theta= 3$, violtaing the bound $\theta \leq d-1$, based on the asymptotic scaling of the entanglement entropy expected to hold in all local quantum field theories \cite{Huijse:2011ef}.}$n(\tilde{\mu})\sim \tilde{\mu}^3+ c \tilde{\mu} $. We find that the value for $c$ is non-zero with great confidence\footnote{Study of subsequent terms in the $\tilde{\mu}$ expansion should reveal contributions from 
bosonic degrees of freedom. However, the scenario of having {\it only} bosonic contributions is excluded.}, as $c=0$ (the value required for hyperscaling to hold) is about 3 sigma away from the central value for the fit for $c$. This fits the expectation of hyperscaling violation exponent $\theta=2$, the value corresponding to small excitations of a Fermi surface.

\begin{figure}[t]
   \centering
    \vspace{-50pt}
    \includegraphics[width=0.35 \textwidth]{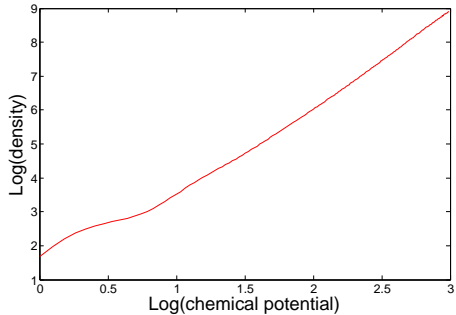}
     \vspace{-50pt}
    \caption{Interface density (normalized to unity at $\tilde{\mu}=1$) as function of the chemical potential, on a log-log scale. The power law behavior of $n(\tilde{\mu})$ at large $\tilde{\mu}$ is apparent, as are deviations from this behavior at lower values of  $\tilde{\mu}$.}

\end{figure}

In summary, we have examined the density per unit volume induced on the interface, revealing structure consistent with the fermionic nature of the boundary excitations. Unfortunately, the densities per unit area cannot be reliably computed using our solutions. Tracking the change in the boundary layer width as function of the parameter  $\tilde{\mu}$ is necessary for that purpose. However, in our appraoch the width of the domanin wall is determined by $a$ rather than by dynamical properties of the interface. It would be interesting to improve on our discussion and calculate reliably the densities per unit area; we leave that for future work.

\section{Conclusions and Outlook}

In this note we constructed inhomogeneous solutions describing the holographic dual of an interface between a fractional topological insulator and an ordinary insulator. We have used these solutions to start the investigation of the gapless strongly correlated fermions expected to live on that interface. Indeed, by studying the low temperature expansion of the thermodynamics of the system we found results consistent with these expectations.

A few directions for further study are immediately apparent. It is interesting to study more general models with fermions bound to an interface, for example models with a  non-trivial dynamical exponent $z$, or models with fermionic excitations only -- though bosons are not expected to form gapless bound states at the interface, they are likely to limit the precision of the analysis. It is also desirable to study such systems at zero temperature and finite density. In that limit the effects of a degenerate Fermi gas are likely to be more pronounced and can be more cleanly studied.  As mentioned in the text, moving away from the "thick" domain wall regime is an interesting, albeit challenging, problem.

Most interestingly, since the boundary fermions are expected to be strongly correlated (i.e. coupled to massless gauge fields), these types of systems are natural candidates for an holographic dual for a non-Fermi liquid. It is therefore particularly interesting to search for further signatures of the fermionic nature of the interface, for example oscillations in response functions, or non-analyticity in the fermionic spectral densities of the system. We expect that general enough models will exhibit characteristics of strange metallic behavior, and we hope to return to some of these questions in the near future.

Finally, beyond that particular example of an holographic interface, there is a whole range of fascinating interface physics to be explored holographically. We expect that the methods used here will be useful in exploring this research direction.

\acknowledgments
I am supported by a discovery grant from NSERC of Canada. I have benefitted from conversations with Marcel Franz, Jeremy Heyl, Gordon Semenoff, Darren Smyth, Jared Stang and Evgeny Sorkin and especially from a correspondence with Kristan Jensen and Andreas Karch, for which I am grateful.

\end{document}